
\input phyzzx
\def\SCIPP{\centerline {\it Santa Cruz Institute for Particle Physics}
  \centerline{\it University of California, Santa Cruz, CA 95064}}
\def\vev#1{\left\langle #1 \right\rangle}
\def\CO{{\cal O}}
\def\diag{{\rm diag}}
\def\abs#1{\left| #1\right|}
\def\ltap{\ \raise.3ex\hbox{$<$\kern-.75em\lower1ex\hbox{$\sim$}}\ }
\def\gtap{\ \raise.3ex\hbox{$>$\kern-.75em\lower1ex\hbox{$\sim$}}\ }
\def\frac#1#2{{\textstyle{#1\over #2}}}
\def\np#1#2#3{{Nucl. Phys. } {\bf B{#1}} (#2) #3}
\def\pl#1#2#3{{Phys. Lett. } {\bf {#1}B} (#2) #3}

\overfullrule 0pt
\titlepage
\date{February, 1993}
\pubtype{ T}     
\title{{Dynamical Supersymmetry Breaking at Low Energies}
\foot{Work supported in part by the U.S. Department of Energy.}}
\author{Michael Dine}
\address{}
\SCIPP
\vskip.2cm
\author{Ann E. Nelson}
\address{}
\medskip
{\vbox{\it\centerline{Department of Physics, 9500 Gilman Drive 0319}
\centerline{University of California, San Diego, La Jolla, CA 92093-0319}
}
\par
\vskip1.6cm
\vbox{
\centerline{\bf Abstract}
Conventional approaches to supersymmetric model building suffer from several
naturalness problems:  they do not explain the large
hierarchy between the weak scale and the Planck mass, and they require fine
tuning to avoid large flavor changing neutral currents and particle electric
dipole moments.  The existence of  models with dynamical supersymmetry
breaking, which can explain the hierarchy, has been known for some time, but
efforts to build such models have suffered from unwanted axions and
difficulties with asymptotic freedom.   In this paper, we describe an approach
to model building with supersymmetry broken at comparatively low energies which
solves these problems, and give a realistic example. }
 \bigskip\submit{Physical Review D}
\vfill
\noindent{SCIPP 93/03, UCSD/PTH 93-05}
\endpage

\parskip 0pt
\parindent 25pt
\overfullrule=0pt
\baselineskip=18pt
\tolerance 3500

\endpage
\pagenumber=1

\chapter{Introduction}

In recent years, supersymmetry has emerged as a leading
candidate for the origin of electroweak symmetry breaking.
Supersymmetry offers a cure to the hierarchy problem (or
at least to the problem of quadratic divergences)
and, in a desert scenario, yields a successful unification
of coupling constants.  Extensive effort has been devoted
to the phenomenology of what has become known as the
``minimal supersymmetric standard model'' (MSSM).  This model has
no difficulty accomodating present experimental constraints.
The features of the MSSM are often understood, or at least
motivated, by considering $N=1$ supergravity theories.

\REF\vadim{M. Cvetic, A. Font, L.E. Iba\~nez, D. L\"ust, and F. Quevado,
\np{361}{1991}{194};  L.E. Iba\~nez, and D. L\"ust, \np{382}{1992}{305}; J.
Louis and V. Kaplunovsky, Texas preprint UTTG-05-93 (1993).} Yet there are a
number of reasons to be concerned about this rosy picture; these concerns can
also
be motivated within the framework of supergravity theories.  First, the
supersymmetry-breaking scale is put in by hand.  The promise of supersymmetry
to
{\it explain} the hierarchy is unfulfilled. Furthermore, the masses of
supersymmetric partners are not calculable, and must also be put in by hand
to be above the experimental bounds. Moreover, in these $N=1$ models, the new
physics associated with the ``hidden sector'' responsible for supersymmetry
breaking is completely inaccessible.  This last objection is  not
fundamental, but it is a disappointing feature of these theories.
Finally, and perhaps most seriously,  there is the ``flavor problem''.
\REF\flavorprob{L.J. Hall, V. A. Kostelecky, S. Raby, Nucl. Phys. {\bf B267}
(1986) 415; H. Georgi, Phys. Lett. {\bf 169B} (1986) 231.}  The pattern of soft
breakings in these theories is highly restricted by flavor changing neutral
currents. Usually it is simply assumed that at some high energy scale, the
squarks and sleptons are degenerate. In the framework of standard supergravity
theories, however, there is no reason for such relations to hold, and indeed
they do not in generic superstring
compactifications.\refmark{\vadim}
More generally, it is hard to understand how theories in which
the quark and squark masses are generated at some very
high energy scale can give rise to significant  squark and slepton
degeneracy\refmark{\flavorprob}. There have been many speculations
on solutions to
all of these problems; still the picture is not completely
satisfying theoretically, and most of these speculations involve physics which
is experimentally out of reach for the foreseeable future.

In the present paper, we wish to consider an alternative picture
of supersymmetry breaking, which has not been considered since
the earliest days of supersymmetry model building.  We wish
to explore the possibility that supersymmetry is dynamically
broken, by new physics associated with (multi) TeV energies.
We will  construct a such a model, where the squark and slepton masses
are calculable.  The model has many desirable features;  in fact,
it will solve all the problems listed above.  It will also make
some predictions, not only about the size of soft breaking terms,
but also about the particle content at the weak scale.  In particular,
beyond the particles of the MSSM,
it predicts the existence of at least one singlet and a set of mirror
quarks and leptons, all at experimentally accessible energies.

\REF\dsb{I. Affleck, M. Dine and N. Seiberg, Phys. Rev. Lett.
{\bf 52} (1984) 1677; Y. Meurice and G. Veneziano, \pl{141}{1984}{69}.}
\REF\dsbphen{I. Affleck, M. Dine and N. Seiberg, Nucl. Phys. {\bf B256}
(1985) 557.}
\REF\acw{L. Ibanez and G. Ross, \pl{110}{1982}{215}; L. Alvarez-Gaume, M.
Claudson
and M.B. Wise, Nucl. Phys. {\bf B207} (1982) 96.}
Models exhibiting dynamical supersymmetry breaking (DSB)
have been known
for some time \refmark{\dsb}.  The authors of ref. \dsbphen\
attempted to construct models with supersymmetry   broken
in the multi-$TeV$ energy range, but ran into a variety
of problems.  Two problems, in particular, seemed generic:
there were light Goldstone bosons, or axions, and QCD
was not asymptotically free, hitting its Landau singularity
a few decades above the scale of the new, supersymmetry-breaking physics.
In the present work, we will exhibit a model which solves
both problems.  The would-be axion will gain mass as a result
of another strong group besides QCD, and the model will
be structured so that QCD is nearly asymptotically flat.  Ordinary
particles -- quarks, leptons, gauge bosons and their superpartners,
``learn of'' supersymmetry-breaking through gauge interactions.
As a result, there is automatically sufficient degeneracy in
the  squark and slepton spectrum to insure adequate suppression of
flavor-changing neutral currents.  $SU(2) \times U(1)$ breaking will arise
through loop corrections to Higgs boson masses involving top quarks,
in a manner discussed long ago \refmark{\acw}.  Breaking  the electroweak
symmetry
without fine tuning will also require the presence of a light $SU(2) \times
U(1)$
singlet, and new superfields with vector-like interactions.

The model we will present here is meant as an existence proof.
It has certain drawbacks.  None of these are fatal, nor is it
clear that  any  of them are generic.  Indeed, we strongly
suspect that a more elegant model is   lurking somewhere.
Perhaps the most serious problem is just that the model is
rather complicated, involving four additional  gauge groups.
It is not, however, nearly as complicated as recent proposals
for technicolor models, and the symmetry group is not larger
than some encountered in string compactifications.  Moreover,
unlike technicolor models, it is not necessary for large numbers
of groups to become strong within a few decades of one another.
Apart from the group actually responsible for breaking supersymmetry,
there is one other group which cannot be too weak; otherwise there
is a light Goldstone boson which is inconsistent with the red giant
and supernova limits.
We will see that at high energies this requires at most a very
mild  fine-tuning.  There is also the potential for generating a large
Fayet-Iliopoulos D-term for hypercharge.  Solving this will require an
approximate equality of certain gauge couplings.  This equality will be seen
to be ``natural" in the sense that it does not receive large radiative
corrections.  Such an equality could arise within the framework of grand
unification or superstring theory. Finally, there are potential cosmological
problems which could be solved by higher dimension operators:  domain
walls, and   long-lived, massive states.  Still, the model has virtues:  a
hierarchy between the weak scale and the shortest distance scales is naturally
generated,  flavor-changing processes  and new sources of CP violation are
naturally suppressed, and it is otherwise consistent with all present day
experiments. Of course, the cosmological constant problem remains as a most
troublesome naturalness issue.

\chapter{Dynamical Supersymmetry Breaking}
 \REF\wittendsb{E. Witten, Nucl. Phys. {\bf B188} (1981) 513.}
\REF\constraints{E. Witten, Nucl. Phys. {\bf B202} (1982) 253.}
The hierarchy problem has two aspects, both of which one might hope
to address within the framework of supersymmetry.  One is the problem
of quadratic divergences of scalar masses.  The second is the existence
of an extremely small dimensionless number, which we can think of
as the ratio of the weak scale to $M_P$ or some unification scale.
It was Witten who first clearly stated in what sense supersymmetry might
solve this second problem.\refmark{\wittendsb}  He noted that
(in the case of global supersymmetry) any small vacuum energy
signifies supersymmetry breaking.  Yet if supersymmetry is unbroken
at the classical level, it remains unbroken to all orders in perturbation
theory as a consequence of non-renormalization theorems.  However, the
proof of this statement is
inherently perturbative, and the result need not hold beyond
perturbation theory. Thus effects of order $e^{-a/g^2}$, where $g$ is some
coupling constant, might give rise to supersymmetry breaking and explain the
large hierarchy.  Witten also formulated
a  set of conditions under which supersymmetry breaking might  or might
not  occur.  Most important of these was the existence (in perturbation theory)
of a massless fermion which could play the role of a Goldstone fermion.
He also showed that a certain index (the ``Witten index") must vanish
if supersymmetry is to be broken, and showed this index to be non-zero
in a number of interesting cases.\refmark{\constraints}

\REF\dsbqcd{I. Affleck, M. Dine, and N. Seiberg, Nucl. Phys. {\bf B241}
(1984) 493.}
Subsequent work showed that in many cases, non-perturbative effects
do violate the non-renormalization theorems.  In some cases these are
due to instantons and can be calculated explicitly
in a systematic semiclassical expansion; in some cases they can be
understood in terms of other non-perturbative effects, such as gluino
condensation. It will be helpful for what follows to review the results for
``Supersymmetric QCD"\refmark{\dsbqcd}.   For our purposes, this is a theory
with
gauge group $SU(N)$ with $N_f$ chiral multiplets in the $N$ representation,
$Q_f$, and $N_f$ in the $\bar N$ representation, $\bar Q_{\bar f}$.  Before
including a mass term or other superpotential term for the ``quark" fields, the
theory has a non-anomalous $SU(N_f)\times SU(N_f)\times U(1)_V \times U(1)_R$
symmetry, where the last two symmetries are a baryon-number-like transformation
and an $R$ transformation with charges chosen to avoid anomalies.
This model also contains, at the classical level, a large set of degenerate
vacuum states, described by several parameters, referred to
as ``flat directions" of the potential.  These are just directions
in which the $D$-terms vanish.  It is not hard to convince oneself
that up to symmetry transformations, for $N_f < N$ the most
general zero-energy
state is described by the expectation values:
$$Q= \left ( \matrix{a_1 & 0 & \dots \cr
0& a_2 & \dots \cr \dots &0 & a_{N_f} \cr
0 & \dots & 0 \cr
\dots & \dots & \dots \cr
0 & \dots & 0 } \right ) =
 \bar Q \eqn\qcdvevs$$

The case where $N_f=N-1$ is the easiest to analyze.  In this case, in these
flat directions, the gauge symmetry is completely broken.  Moreover, by
choosing the vev's large enough, one can make the theory arbitrarily
weakly coupled (the effective coupling is $\alpha(M_V)$, the asymptotically
free coupling of the theory evaluated at the scale of the gauge boson
masses).  In these vacua the light degrees of freedom are a set of Goldstone
bosons and their superpartners.  These can be described by a matrix-valued
field $\Phi_{f,\bar f}= Q_f \bar Q_{\bar f}$.  In these vacua, in order to
determine if supersymmetry breaking occurs, one must compute the effective
action for these light degrees of freedom.  In particular, supersymmetry
breaking requires that one generate a superpotential for these
fields.  Any superpotential must respect the original flavor symmetries of the
theory. The $SU(N_f)\times SU(N_f)$ symmetry implies that the action
must be a
function only of $\det(\Phi)$.  The $U(1)_R$ symmetry then determines the form
of the superpotential uniquely:
$$W = {\Lambda^{2N+1}\over \det(\Phi)}\eqn\wforn1$$
where $\Lambda$ is the scale parameter of the $SU(N)$ group.
A completely straightforward instanton calculation yields precisely
the various component field interactions implied by this
lagrangian\refmark{\dsbqcd}.

For $N_f < N-1$, one can repeat most of the analysis above.  In particular,
the symmetries determine the form of any superpotential uniquely,
in terms of the $N_f \times N_f$ field,
$\Phi_{f,\bar f}= Q_f \bar Q_{\bar f}$:
$$W = b\Lambda^{3N-N_f \over
N-N_f} \det(\Phi)^{-1/(N-N_f)}\eqn\wgeneral$$
A somewhat different analysis is required to show that this superpotential
is generated in this case.  For $N_f > N-1$, no superpotential is generated.

What are the implications of this superpotential?
These examples illustrate that the non-renormalization
theorems {\it do} break down non-perturbatively.  However, at least at weak
coupling, these theories don't have a good ground state, and at best admit
a cosmological interpretation.  If one adds mass terms for the quarks,
the potential is stabilized, and one finds $N$ supersymmetric ground
states, in agreement with Witten's calculation of the index.  If one
wants to find theories in which supersymmetry is broken and
one has a good ground state, it is necessary to study chiral theories.
In ref. \dsbphen, a number of chiral theories were studied which do exhibit
DSB.   The main conditions for this
are:  1)  the absence of flat directions in the classical theory
2)  the existence of a non-anomalous, continuous symmetry
which is spontaneously broken.  These conditions are not hard
to understand.  The second implies the existence of a Goldstone boson.
Unbroken supersymmetry would imply the existence of a  {\it scalar} partner
for this field, but this would impliy the
existence of a flat direction, contradicting  (1).  One can imagine
loopholes to this argument, but these conditions seem to be a good
guide to finding theories with DSB.

\chapter{Strategies for Model Building}

\REF\macintire{M. Dine and D. Macintire,   Phys. Rev. {\bf
D46} (1992) 2594.} Having established the existence of models with
dynamical supersymmetry breaking, it is natural to try and build realistic
models of low energy supersymmetry incorporating it.  There are two strategies
one might adopt.  First, one might use these models as hidden sectors for
$N=1$ supergravity models.  Aspects of this problem have been discussed
elsewhere .\refmark{\macintire}  However, even if this program is
successful, it has little predictive power; the superparticle spectrum remains
a function of unknown parameters, and the origin of degeneracy
among squarks and sleptons remains mysterious.

Alternatively, one can consider the possibility that supersymmetry is
broken at comparatively low energies, and that the breaking of
supersymmetry is fed to the superpartners of ordinary fields through
gauge interactions.  The most straightforward (though perhaps not the
most clever) way to proceed is to take a model of the type discussed
above with DSB, and gauge a global symmetry, identifying it with one
of the usual gauge interactions.   The squarks, sleptons, and gauginos
will gain mass through loop effects. Previous efforts to realize this scenario
floundered on two problems.  First, there is often a problem with the
asymptotic freedom of QCD.   Models with DSB and large enough global symmetry
groups to gauge an SU(3) subgroup  typically contain large numbers of triplets
and anti-triplets.  For example, in ref. \dsbphen, the simplest  such model had
gauge group $SU(11)$ and gave rise to $11$ new flavors of quarks, while
in a supersymmetric theory  if one requires that QCD not have a Landau pole
below the unification scale  of $10^{16}$ GeV  at most four new flavors of
quarks
are allowed at the weak scale.

A second problem is the existence of unacceptable Goldstone bosons or
axions. As we have noted, all known examples of dynamical supersymmetry
breaking   require the presence of a spontaneously broken global symmetry,
and with it a massless Goldstone boson.  In many cases,
this boson is an axion, once QCD is taken into account.  In any case,
the decay constant of this boson is of multi TeV order, and thus it is
typically inconsistent with astrophysical limits.

Here we will describe a model which avoids both of these problems.
The dangerous Goldstone boson will gain mass as a result of anomalies
with respect to another strong gauge interaction, known as ``R-color''.   The
problem of non-asymptotic freedom will be avoided by using a slightly more
complicated strategy than that described above.
The extra gauged symmetry in the ``supercolor sector" (i.e. the sector
responsible
for breaking supersymmetry) will not be identified with the standard
model gauge interactions, but with R-color.  There will be some
additional fields carrying R-color as well as ordinary
gauge quantum numbers, which will cancel R-color anomalies and act as the
``messengers" of supersymmetry breaking.  As a result, QCD will be only barely
non-asymptotically free.

 One unpleasant feature of the model, which differs from earlier models of
DSB, is that it has classically flat directions. The degeneracy is lifted by
nonperturbative effects, but a supersymmetric minimum appears at infinite
value of some scalar fields; the theory has no ground state.
However we can find
a {\it local} minimum of the potential which violates supersymmetry.  For small
coupling, this minimum will be essentially stable against tunnelling.  We won't
worry here about how the universe might have found itself in this state.

\chapter{The Model}

\section{Fields and Lagrangian}

Let us now turn to the actual model.  Apart from $SU(3)\times SU(2)\times
U(1)$, the gauge group of the model is
$$SU(7)\times SU(2)\times SU(3)_L \times SU(3)_R  \eqn\gaugegroup$$
The $SU(7)\times SU(2)$ groups will be referred to as ``supercolor."
The $SU(7)$ gets strong, and nonperturbative terms in the superpotential
generated by the $SU(7)$, in conjunction with the D-terms from the $SU(2)$,
will
be responsible for supersymmetry breaking.  The matter fields of the model
consist, first, of the usual quark and lepton superfields, a pair of Higgs
doublets, and a singlet, $S$.  The latter particle will be necessary for
achieving $SU(2) \times U(1)$ breaking.  To describe the additional fields of
the model, it is convenient, first, to ignore the standard model fields and
interactions, and impose a global $SU(7)$ symmetry. The usual gauge
interactions will lie within this $SU(7)$.  This procedure will allow us to
turn immediately to the essential dynamical features of the model.  Later we
will return to the realistic situation where the global $SU(7)$ symmetry is
explicitly broken to an $SU(3)\times SU(2)\times U(1)$ gauged subgroup.  Under
$SU(7)\times SU(2)\times SU(3)_L \times SU(3)_R  \times SU(7)_G$, the
additional fields are: $$Q = (7,1,\bar3,1,1,) ~~~~\bar Q = (\bar
7,1,1,3,1)$$ $$q=(7,2,1,1,1)~~~~\bar u=(\bar 7,1,1,1,1)~~~~\bar d=(\bar
7,1,1,1,1)$$ $$X= (1,1,\bar 3, 3, 1)~~~~\bar X= (1,1,3,\bar 3, 1)$$ $$f=
(1,1,3,1,7)~~~~\bar f = (1,1,1,\bar 3, \bar 7)$$
$$l=(1,2,1,1,1)\eqn\quantumnos$$ The superpotential of the model is $$W=
\lambda_1 \bar Q X Q + {\lambda_2 \over 3} \det(X^3) + \lambda_3 q \bar u l+
{\lambda_4 \over 3} \det(\bar X^3) + \lambda_5 \bar f \bar X f \eqn\treew$$
Note that the model is anomaly free.  When we return to consider the ordinary
$SU(3) \times SU(2) \times U(1)$, we will simply imbed this group
in the standard way in an $SU(5)$ subgroup of $SU(7)$.  We will
thus take $f$ and $\bar f$ each to break up into a triplet, a doublet,
and two singlets.  The coupling $\lambda_5$ will then actually
represent four independent parameters, which we refer to as
$\lambda_5^{t,d,s,s'}$.

We will suppose that $SU(7)$ is the strongest group
(i.e. the one with the largest $\Lambda$-parameter), and that all
of the couplings in the superpotential are small.  In this approximation,
the $SU(7)$ sector of the theory is an example of supersymmetric QCD
with seven colors and five flavors.  Grouping
the $7$ and $\bar 7$'s of the theory into fields
${\cal Q}$ and $\overline{\cal Q}$,
this theory has flat directions of the form
of eqn. \qcdvevs,
(it is necessary to use the approximate $SU(5) \times SU(5)$ flavor
symmetry to bring these fields to this form).  Non-perturbatively
as described above, a superpotential
of the form
$$W_{np}= {\Lambda^8 \over (\det\Phi)^{1/2}}\eqn\susevenw$$
 is generated, where
$$\Phi = ({\overline{\cal Q}}_{\bar f} {\cal Q}_f)\eqn\phidef\ .$$

For small couplings $\lambda_i$, we want to study the potential
$$W=W_{cl}+W_{np}\eqn\totalw$$
(and the $D$-terms for the various groups).
We expect that for small $\lambda_i$,
the minimum of the potential will
lie at large values
of the fields, justifying this analysis.  Provided $\lambda_i \ll g_a$
(the gauge couplings), we should be able to find the minima by looking
at flat directions of the $D$-terms.
Of course, the full theory does not have the $SU(5)\times SU(5)$
symmetry used above, and so one must consider a more general set of
flat directions of the $D$-terms.  The structure of the
complete potential is quite complicated, and we will not be able to
survey the entire field space.  Instead, we will look for a local minimum
of the potential in a particular direction in the field space.
This will be described by the expectation values:
$$Q= \left ( \matrix{a & 0&0 \cr 0& a & 0 \cr 0& 0& a \cr 0 & \dots & 0 \cr
\dots & \dots & \dots } \right )
= \bar Q~~~~X=\diag(x,x,x)$$
$$q= \left ( \matrix{0 & 0 \cr 0 & 0 \cr 0 & 0 \cr b & 0 \cr 0 & c \cr
0 & 0 \cr 0 & 0} \right ) ~~~
\bar u = \left ( \matrix{ 0 \cr 0 \cr 0 \cr d \cr e \cr 0 \cr 0  }\right )
  \bar d= \left ( \matrix{ 0 \cr 0 \cr 0 \cr f \cr g \cr 0 \cr 0  }
\right )
L=\left ( \matrix{ h \cr 0} \right ) \eqn\ourdirection$$
with
$$\vert f \vert^2 = \vert b \vert^2 -\vert d \vert^2
{}~~~~~ \vert g\vert^2 = \vert c \vert^2 -\vert e \vert^2
{}~~~~~\vert h \vert^2 = \vert c \vert^2 -\vert b \vert^2\ .
\eqn\dtermconditons$$
We will establish that there exists a {\it local} minimum
of the potential of this form.  Note that this minimum leaves over an
$SU(3)$ gauge symmetry which is a linear combination of the
$SU(3)_L$, $SU(3)_R$ and an $SU(3)$ subgroup of $SU(7)$. This $SU(3)$, known
as R-color, will subsequently also get strong. There is also an unbroken
$SU(2)$ subgroup of the $SU(7)$, which will play no role in the subsequent
discussion.
The fields $f$, $\bar f$, and $\bar X$ play no role in the dynamics which
break supersymmetry; but simply serve as ``messenger'' particles which
communicate supersymmetry breaking to    ordinary  superfields.

The rest of the model, which we will refer to as the ordinary sector, just
consists of the MSSM, without the bilinear $\mu H_1 H_2$ term in the
superpotential (this can be eliminated by imposing a $Z_3$ discrete
symmetry). We get rid of this term because otherwise electroweak symmetry
breaking would require an unacceptable fine-tuning of $\mu$.
Instead, breaking $SU(2)\times U(1)$ will require the addition of a
gauge singlet $S$  and   vector-like superfields  $D,\bar D, L, \bar L$
transforming as $(3, 1, -1/3),(\bar 3,1,1/3),(2 ,1 , 1/2),(2,1,-1/2)$ under the
ordinary $SU(3)\times SU(2)\times U(1)$ interactions.

\FIG\oneloopmasses{One loop diagrams contributing
to the masses of the $\bar X$, $f$ and $\bar f$ fields.}

\section{Overview of Symmetry-Breaking in the Model}

Before going through the detailed analysis of the model, we summarize   the
basic features.  Having established that the minimum of the potential is of the
form of eqn. \ourdirection, one of our main goals is to determine the masses
of squarks, sleptons and gauginos, as well as Higgs particles.  These will
arise as a result of the gauge couplings of the $f$ and $\bar f$ fields.
The scalar components of these fields, as well as the fields $\bar X$, can gain
mass at one loop through graphs such as those shown in fig. \oneloopmasses.
It turns out, that for a range of parameters, the mass-squared's of these
fields are negative.  Minimizing the resulting potential yields
a vev for $\bar X$ of the form
$$\bar X = \diag(\bar x,\bar x,\bar x)~~~f=\bar f = 0\eqn\xbarmin$$
The fields $f$ and $\bar f$ receive an additional contribution to their
mass from the vev of $\bar X$.  Below the scale of these vev's, one
has an unbroken $SU(3)$ gauge theory without matter fields at all
(the $SU(3)$ gauginos gain mass at one loop).

\FIG\gauginomass{One loop diagrams contributing to the masses of the
gauginos.}
\FIG\squarkmass{Two loop diagrams contributing to squark, slepton
and Higgs masses.}
\FIG\topquarkloop{Loop correction to Higgs mass in the low energy effective
theory, which gives negative contribution proportional to the squark mass
squared.} Ordinary gauginos gain mass through diagrams of the type shown in
fig.
\gauginomass.  Squarks and sleptons gain masses through diagrams of the type
shown in fig. \squarkmass. For a range of parameters, these contributions can
be
shown to be positive. Note that the masses of squarks, sleptons and gauginos
depend in a simple way on their gauge couplings.  Squarks are generically
heaviest, lepton doublets and Higgs are lighter by roughly a factor $\alpha_2
/\alpha_3$, and singlet leptons are lightest.
In order that $SU(2)_L$ be broken, it is necessary that one Higgs particle
obtain a negative mass-squared.  This
can occur for the Higgs which couples to top quarks \refmark{\acw}, as a result
of the
diagram of fig. \topquarkloop.  While this diagram is nominally one
higher order in the loop expansion, it is enhanced by the fact that
the squark masses are larger than the doublet masses by a factor
$\alpha_3 \over \alpha_2$, and by a logarithm, and can be larger than
the positive two loop contributions.  For a range of parameters, as a result,
the Higgs mass-squared can be negative. In this model, however,  there is no
$H_1
H_2$ term in the potential, so in order that there be a suitable quartic
coupling
for the Higgs field, it is necessary to include a singlet field, with couplings
$S H_1 H_2$ and $S^3$. Furthermore, in order to obtain a sensible breaking of
$SU(2) \times U(1)$ with masses for all the quarks and leptons and without fine
tuning it will be necessary to add new vector-like fermions carrying
$SU(3)\times
SU(2) \times U(1)$ gauge charges, which couple to the singlet and gain mass
from
its vev.

Finally, we have to worry about the various global symmetries of the
model.  The vector-like symmetries are preserved at this minimum,
but the $U(1)_R$ symmetry is broken, giving rise to an axion.   This symmetry
has
no $SU(7)$ anomaly. However the   axion does  get a mass from
the unbroken $SU(3)$ R-color, which is of order the scale at
which this group gets strong, squared, divided by its decay constant.  This
mass
can easily be of order $10~MeV$, so its production in stars can be adequately
suppressed.  There remain a number of cosmological worries about this model;
these include domain walls and stable massive particles, and will be dealt with
later, as will the question of Fayet-Iliopoulos terms.  We now turn to a
detailed
discussion of each of these points.

\section{Supersymmetry Breaking}

First, let us turn to the problem of minimizing the potential.
It is not hard to see that the minimum in the
direction of eqn. \ourdirection\ cannot be the global minimum;
the global minimum has zero energy.  At the classical level
the theory has a flat direction with
$$X= \diag(x,0,0) = \bar X\eqn\flatdirection$$
all other fields vanishing.
Once one considers the non-perturbative piece of the superpotential,
this direction is no longer flat.  However, it is possible to
let $X\rightarrow \infty$, $Q\rightarrow \infty$,
and the fields $q$, $\bar u$, $\bar d$ and $L$ tend to zero
in such a way that the total energy tends to zero.  Note first that
for large $x$, the unbroken symmetry is an $SU(2) \times SU(2)\times
U(1)$.   One $Q$ flavor gains mass, as does a $(2,2)$ field from $X$.
After integrating out massive states, there is no dimension four
term in the superpotential for the light $Q$'s.  In order to minimize
the non-perturbative contribution to the potential, then, one wants
to let $Q$ get large (though not as fast as $X$), while the other
fields get smaller more slowly.  For example, the scaling
$$Q \sim x^{1/4}\sim \bar Q~~~~ q \sim x^{-1/8}\sim \bar u \sim\bar d$$
gives an energy tending to zero as $x^{-1/2}$.
As explained in the introduction, we will not worry about the global
structure of the potential, and simply assume that somehow the
universe finds itself in the vacuum of interest, and does not tunnel out.

Let us turn to the problem of minimizing the potential in the
direction of eqn. \ourdirection.  The problem is easiest to
analyze in the limit $\lambda_3 \ll\lambda_1  \ll \lambda_2$.
In the limit $\lambda_3 \rightarrow 0$, the auxiliary fields, $F_X$ and
$F_Q$ should vanish.  $F_X=0$ gives
$$x^2= \left({\lambda_1 \over \lambda_2}\right)a^2\eqn\xsolution$$
while $F_Q=0$ gives an expression for $a$ in terms of $\det\Phi$:
$$a= {1 \over 2} \lambda_2^{1/2} \lambda_1^{-3/2}\Lambda^8 \tilde
\Phi^{-1/2}\eqn\asoln$$
where we have defined
$$ a^4\det\Phi =\tilde \Phi\eqn\phitilde$$
Now one can plug this expression for $a$ into the remaining terms in the
potential:
$$\vert {\partial W \over \partial q} \vert^2
+\vert {\partial W \over \partial \bar u} \vert^2
+\vert {\partial W \over \partial \bar d} \vert^2
+\vert {\partial W \over \partial l} \vert^2\eqn\threepot$$
We can obtain the dependence of the vev's $b-h$
on the couplings $\lambda_1 \dots \lambda_3$ by scaling arguments.
A simple exercise gives
$$(b,c,d,e,f,g,h) \sim \lambda_3^{-1/4} \lambda_1^{3/16} \lambda_2^{-1/16}
\Lambda\eqn\bscalings$$
One can check that, for a finite range of parameters, the minimum of the
potential is indeed of this form, with an unbroken $SU(3)$.

For finite but small $\lambda_3$, the relations
$$F_Q = 0~~~~F_X=0\eqn\fqfx$$
are not exactly satisfied.  To determine the corrections, we need to compute
the shifts in the vev's $a$ and $x$ to the next non-trivial order in
$\lambda_3$.  Again, it is not hard to determine how these scale with
couplings.  The shift in $a$, $\delta a$, can be determined by
computing the $a$ tadpole, $\partial V \over \partial a$, and dividing
by the $a$ mass-squared.  Using our scaling results above, one finds that
$$\delta a \sim\Lambda  \lambda_2^{7/16}\lambda_3^{3/4}\lambda_1^{-21/16}$$
$\delta x$ is smaller by a factor $\sqrt{\lambda_1\over\lambda_2}$.  We can
estimate $F_X$ and $F_Q$ by writing: $$F_X = {\partial^2 W \over \partial a
\partial x} \delta a + {\partial^2 W \over \partial x^2} \delta x\eqn\fx$$
with a similar equation for $F_Q$.  Note that the second derivatives
here are just elements of the lowest order mass matrix.  In the
limit $\lambda_1 \ll \lambda_2$, one finds that $F_X$ is largest,
$$F_X    \sim \Lambda^2 \lambda_2^{13/ 24}\lambda_3^{5/
6}\lambda_1^{-5/ 8}\gg F_Q \sim \Lambda^2
\lambda_2^{1 / 24}\lambda_3^{5/ 6}\lambda_1^{-1/ 8}\eqn\fxestimate$$

Note that the spectrum of particles in the supercolor sector is
nearly supersymmetric.  The breaking of supersymmetry is
represented by the small values of the $F$-components
(small by powers of the couplings in the superpotential),
which give rise to small splittings within the multiplets.
The gauge bosons associated with the broken $SU(3)_L \times
SU(3)_R$ are also nearly supersymmetric.  The spectrum
is  simpler to work out if the gauge couplings of these
groups, $g_L$ and $g_R$ are identical; as we have already
remarked in the introduction, this condition must in fact
be satisfied if the model is to be realistic.  The expectation
value $a$ is larger than that of $x$, so, neglecting the $x$ vev
there are two massive gauge bosons, with mass squared
$g_L^2 a^2$, and $(2 g_7^2 + 2 g_7 g_L) a^2$, and one massless
eigenstate corresponding to the unbroken $SU(3)$.
If we assume $g_L \ll g_7$, the former is the lighter
state; it is simply the linear combination
$$B^{\mu} = {1 \over \sqrt{2}} (A_L^{\mu} + A_R^{\mu})\eqn\lightstate$$

\REF\ovrut{B. Ovrut and J. Wess, Phys. Rev. {\bf D 25} (1982) 409.}}
\FIG\supergraphs{Supergraphs contributing to the
scalar masses.  Solid lines are chiral superfields; wavy
lines denote gauge fields.  X's denote vacuum insertions.}
\FIG\dropfig{Examples of diagrams suppressed by powers of couplings.}
This hierarchy of vev's will be important when we estimate
the loop contributions to various masses.
At the classical level there are many massless states, such as $f$,
$\bar f$ and $\bar X$.  To determine whether these fields obtain
expectation values, one needs to compute their masses.
These will arise from the one-loop diagrams shown in fig. \oneloopmasses.
In the limit in which we are working, in which supersymmetry-breaking
is small, one can evaluate the masses perturbatively in powers of
$F_X$.  This is conveniently done using supergraph techniques.
The required diagrams are then indicated in fig. \supergraphs.
Because of the hierarchy of vev's, it is not necessary to consider
diagrams such as that of fig. \dropfig, with external $X$'s; it is also
not necessary to consider diagrams with external $F_Q$'s.
To evaluate the diagrams it is convenient to chose the supersymmetric
analog of $R_{\xi}$
gauge.\refmark{\ovrut}  In this gauge, the gauge propagator
is simply
$$\Delta= {\delta^4(\theta_1-\theta_2) \over p^2-M_V^2}\eqn\gaugeprop$$
The $\theta$ integrations are trivially performed, and one obtains
for the scalar mass
$$m_S^2 = -{\cal C}{g^4\over 16 \pi^2}{ F_X F_X^{\dagger}  \over
M_V^2}\eqn\ms$$
${\cal C}$ is a group-theory factor which is easy to work out in each case.
For example, under the surviving $SU(3)$, $\bar X$ decomposes as a singlet,
$x^s$ and an octet, $x^a$; for these, ${\cal C}$ equals $8/3$ and $7/6$,
respectively.  For the $f$ and $\bar f$ fields, all of which are triplets,
${\cal C}= 2/3$.

We wish to determine the pattern of symmetry breaking at this stage.
In particular, we will ask if the effective potential has a local minimum
at which $SU(3)$ remains unbroken; this requires that only the
singlet, $x^s$, obtain a vev.  To investigate this, we need to
determine the form of the quartic terms in the potential, which arise
from two sources.  First, there are the terms in the original superpotential.
In terms of canonically normalized fields, this superpotential takes the
form
$$W = {\lambda_4 \over  \sqrt{3}}({x^{s3}\over 3} -{1\over 2}x^s x^{a2}
+{\cal O}(x^{a3}))\ ,\eqn\suthreew$$
where
$$\bar X ={ x^s\over {\sqrt{3}} }+\sqrt{2} x^a T^a\eqn\barxdecomp\ .$$
If this were the end of the story, it is easy to check that the
$SU(3)$-preserving extremum of the potential (including
the loop-generated mass terms) is unstable.  If one simply
looks for an extremum with $x^s \ne 0$, $x^a=0$, one finds
that the octet masses are tachyonic.

\FIG\dcancellation{In the supersymmetric limit, $D$ terms
in the potential associated with broken gauge generators are
cancelled by exchange of the massive scalars in the vector multiplet.}
However this is {\it not} the whole story; there are additional tree level
supersymmetry breaking quartic couplings in the effective low energy theory
which describes the $\bar X$ fields.  To understand this, consider the terms in
the potential of the full theory associated with the auxiliary $D$ fields
for $SU(3)_L$ and $SU(3)_R$.  These terms   are non-vanishing for
$\bar X$ fields of the type we are describing (remember that
$\bar X$ transforms as a $(\bar 3, 3)$).  If supersymmetry
were unbroken, this would be irrelevant at low energies.  Integrating
out the massive gauge multiplet, these $D$ terms would not appear
(corresponding to the fact that effects of small vev's for $\bar X$ would
be cancelled by shifts of the massive fields).  The cancellation of the
$D$ terms would arise from the diagram of figure \dcancellation.
In this diagram, the exchange of the massive scalar in the gauge
multiplet (one of the superpartners of the massive gauge boson)
precisely cancels the quartic couplings associated with the $D$-terms.

When supersymmetry is broken, however, this cancellation is not
complete.  The scalars in the multiplet are no longer exactly
degenerate with the
gauge bosons.  As a result, there is a quartic coupling remaining in
the low energy theory.  Even without a detailed computation, it is
easy to determine the sign and order of magnitude of this coupling.
Suppose, first that the gauge coupling were zero.  Assuming, as we
have above, that the $F$ terms all have small vev's, there
will be a nearly degenerate Goldstone supermultiplet, consisting of
a Goldstone boson, a light fermion, and an additional scalar particle.
This scalar will have a positive mass-squared of order $\abs{
\lambda \vev F }^2$, where $\lambda$ is the coupling to the Goldstino;
this is a consequence of a famous sum rule.  It is easy to check this
in simple examples.  In the present case, this leads to a positive
quartic coupling of the $\bar X$ fields.

Our remarks above can be summarized by the statement that, in
addition to the terms in the potential arising from the above
superpotential, the potential contains the supersymmetry-breaking
terms
$$V_{soft}=-m_s^2 \vert x^s\vert^2 -m_a^2 \vert x^a\vert^2
+{ g_L^2\over 2} \gamma \vert x^s\vert^2 \vert x^a\vert^2$$
where the last term arises from the incomplete cancellation of the
$SU(3)_L$ and $SU(3)_R$ $D$-terms, and is of order $\vert F
\vert^2 \over M^4$, where $F$ is a typical $F$-term in the supercolor
sector, while $M$ is a typical mass.  It is easy to see that, for a range
of parameters, the potential has a local minimum at which
$$x^{s2} = {3 m_s^2 \over \lambda_4^2}~~~~~~x^a=0
{}~~~~~f=\bar f = 0\ . \eqn\barxmin$$

What does the theory look like at this minimum?  The $SU(3)$
symmetry is still unbroken, but, due to the vev of $x^s$, all
of the fields which carry $SU(3)$ quantum numbers gain mass.
(Note that the  gaugino of the $SU(3)$ gains mass also at one loop.
\foot{Note that, because both the
scalar {\it squared}-masses and the gaugino
masses arise at one loop, the gauginos
are generically lighter than the $f$, $\bar f$ and $x^a$ fields,
and will contribute to the renormalization group evolution for
a decade or so.})  Thus we have an effective pure $SU(3)$ gauge theory.
This R-color theory   is now quite asymptotically free, and, depending on
the precise values of the $SU(3)_L$ and $SU(3)_R$ couplings, can get
strong rather quickly.

The asymptotic freedom of R-color is phenomenologically
essential.   It gets rid
of a Goldstone boson produced by the supercolor interactions. As we have noted
earlier, the theory posseses an $R$ symmetry, explicitly broken by
$SU(3)_L$ and $SU(3)_R$ anomalies.  This symmetry has an ordinary color
anomaly and is spontaneously broken by the supercolor sector, giving rise to
a not very invisible  axion with   couplings to ordinary quarks.  In order to
be consistent with astrophysical bounds, this pseudo Goldstone boson must gain
a mass of order a few MeV, at least. The mass of the axion is
on the order of $\Lambda_R^2/\Lambda_7$. Since, as we will see, supersymmetry
breaking and $SU(2)\times U(1)$ breaking in the ordinary sector are only
achieved at two and more loops, we must have the supercolor scale be rather
large compared with the weak scale; $\Lambda_7\sim 10^7$ GeV. Thus the scale
$\Lambda_R$ of R-color   must be greater than  $\CO(100)$ GeV; this in turn
requires that the $SU(3)_{L,R}$ couplings be rather large, but   there is a
finite range of $g_L, g_R$ for which this condition is satisfied, and yet the
couplings do  not blow up below $M_p$.

\section{The ``Low Energy" Spectrum}

We now wish to ask about the spectrum of ``ordinary" squarks,
sleptons, and gauginos.  We will see that, in the effective theory
below the scale of the $f$, $\bar f$, and $\bar X$ fields,
the gauginos gain mass at one loop, while squarks and sleptons
gain positive mass-squared at two loop order.  The problem of
$SU(2) \times U(1)$ breaking will be taken up in the next section.

\FIG\fayet{One loop diagram contributing to a Fayet-Iliopoulos
term in the effective theory.}
We first have to address another potential
problem in the model:  the appearance of a Fayet-Iliopoulos
$D$-term for hypercharge as we integrate
out the heavy fields, $f$ and $\bar f$.  Such a term is  phenomenologically
dangerous, since if it is large it could lead to very light squarks and
sleptons, or even   squark   vevs.   Suppose $g_L \ne g_R$, and
$\lambda_5^t\ne\lambda_5^d$.  Then the diagram of fig. \fayet\ leads, in
general, to a non-zero Fayet-Iliopoulos term.   The relation
$\lambda_5^t=\lambda_5^d$ is renormalized at one loop, and so we assume it
does not hold.  Thus it is necessary to insist that $g_L  = g_R$, to a rather
high degree of accuracy (roughly of order $\alpha \over \pi$). If this were the
case, the full theory, ignoring ordinary quarks and leptons, would possess a
left-right symmetry which would insure the absence of a $D$-term.  Of course,
any such symmetry is broken by the gauge couplings of quarks and leptons, so
one must ask how natural the relation $g_L=g_R$ is.  First, note that radiative
corrections to this relation will arise only at high loop order.  Second,
recall that in string theory, one typically has equality of various gauge
couplings at tree level. If that were the case here, the subsequent evolution
of these couplings would induce only small differences in $g_L$ and $g_R$.
Thus it does not seem implausible to make such an assumption.

{}From now on, we will assume that $g_L$ and $g_R$ are equal, and that
any Fayet-Iliopoulos term is very small.  We turn
to the computation of the gaugino and squark and slepton
masses.  Again, we consider first the effective theory below
the supercolor scale.  In this theory, the $f$ and $\bar f$
fields have soft supersymmetry-breaking corrections to
their masses.  If we integrate out these fields, gauginos
will obtain mass at one loop from graphs such as those shown in
fig. \gauginomass.  These will lead to masses of order
$$m_i = {\alpha_i \over \pi} {\vev{F_{\bar X}}
\over \vev{\bar X}}\  .\eqn\gauginomass$$
Note the result that the gaugino masses are proportional to their gauge
couplings  squared, just as in the usual grand unified $N=1$ minimal
supergravity models. This result only holds when
$\vev{F_{\bar X}}$
is small compared with   $\vev{\bar
X}$. Otherwise  gaugino masses depend on $\lambda_5^t$ and  $\lambda_5^d$; and
for $\lambda_5^t \ne \lambda_5^d$  need not satisfy the GUT
relations.

Squark and slepton masses will arise at two loops from the diagrams
shown in Fig. \squarkmass.  These diagrams are logarithmically
divergent.  The upper cutoff should be interpreted as the supercolor
scale (if one wants to obtain the subleading terms, it is necessary to
``open up" the mass insertions, computing three-loop diagrams
including supercolor fields).  It is not difficult to compute the
logarithmic term. (In this computation, it is perhaps worth noting that the
separate
diagrams exhibit a $\log^2(\Lambda)$ behavior, but the final
answer only contains a single logarithm.)
One obtains
$$\widetilde m^2 = -{C_F \over 4} \left({\alpha_i \over \pi}\right)^2
\delta m_f^2 \ln(M^2/m_f^2) $$
where $C_F$ is the quadratic Casimir of the matter
representation (e.g. $4/3$ for color triplets, $3/4$ for
SU(2) doublets) and $\delta m_f^2$ is the supersymmetry
breaking mass-shift of the $f$ and $\bar f$ fields; note
that this quantity is negative.

The main features to note about this result are that it is positive
(so color and electric charge can remain unbroken), and that
the scalar masses, in this approximation, depend only on gauge
quantum numbers, so flavor-changing processes are adequately
suppressed. Also, the squark and slepton masses are logarithmically enhanced
compared with the gaugino masses.

In summary, the superpartner spectrum in these models is computable,
although unfortunately it depends on several new coupling constants. However
assuming that
the superpotential couplings $\lambda_5^{t,d}$ are   comparable, we
can make the following rough predictions for the squark , slepton and gaugino
masses: $$\eqalign{
{m_3}/g_3^2\approx &{m_2}/g_2^2\approx (3/5) {m_1}/g_1^2\cr
\widetilde{m}_q\sim & \sqrt{\log(M^2/m_f^2)}{m_3}\cr
\widetilde{m}_l\sim &  (g_2^2/g_3^2)\widetilde{m}_q \cr
\widetilde{m}_e\sim & (g_1^2/g_3^2)\widetilde{m}_q \  ,
\cr}\eqn\roughpredict$$ where $\widetilde{m}_q$ is the mass of the
(nearly degenerate) squarks, $\widetilde{m}_l$ is the mass of the slepton
doublets, and $\widetilde{m}_e$ is the mass of the slepton singlets.
As we will see in the next section, the weak scale is determined by a three
loop negative contribution to the Higgs mass squared, which is comparable to
${\widetilde{m}}_l^2.$ Thus we expect $\widetilde{m}_l\sim v=250$ GeV, which
gives for the approximate size  of the other scalar superpartner masses
$\widetilde{m}_e \sim 100$ GeV, $\widetilde{m}_q\sim  900$ GeV.  The slepton
$SU(2)$ singlets could be within reach of LEP II.

\section{$SU(2)\times U(1)$ Breaking}
In this section we turn to the problem of electroweak symmetry
breaking.  At two loops, we have obtained positive
masses for all of the scalar fields in the low energy theory.
If $SU(2) \times U(1) $ is to break, at least one Higgs field
must acquire a negative mass-squared. For this to happen, a three loop
negative contribution to the mass squared   must be larger
than the two loop contributions.
As pointed out long ago \refmark{\acw}, in a model such as this one it is easy
to
obtain a negative mass-squared for the Higgs which couples to the
top quark.  The point is that the loop corrections of fig.~\topquarkloop, while
suppressed by a factor $3 g_t^2/( 16 \pi^2)$ are enhanced both by a
logarithm of $m_f^2/{\widetilde{m}}_q^2$ and by the fact that the  top squark
mass itself is proportional to $\alpha_s^2$ rather than $\alpha_W^2$, as for
the lowest order Higgs mass.  Thus for top quarks in the presently allowed
mass range, this three loop graph  can give a negative contribution to the
Higgs mass squared which is larger than the two loop positive contribution.  To
see the logarithmic enhancement, it is convenient to study the effective theory
below the mass scale $m_f$ of the messenger particles. In the low energy
theory,  the  graph in fig.~\topquarkloop, which is  proportional to the
large squark mass  squared ${\widetilde m}_q^2$, causes the mass squared
$M_1^2$ of  $H_1$ to run. (There will be other contributions to the
renormalization group equations in the effective theory, \eg from trilinear
scalar terms and gaugino masses, but these are smaller).   $M_1^2$ is positive
at
$m_f$, and decreases rapidly. If $M_1^2$ becomes negative at a scale above
$\widetilde m_q$ then $H_1$ will get a vev. However in order to give masses to
all quarks and leptons, both $H_1 $ and $H_2$ must get vevs.   The symmetries
of
the model prevent the generation of a $m_{12}^2 H_1 H_2$ term in the potential.
Furthermore,  it is
not   easy to obtain a negative mass-squared for $H_2$, since, in general, the
bottom
quark Yukawa coupling, $g_b$, is not as large.   One can, of course, try to
choose couplings so that $\vev{H_2} \ll \vev{H_1}$.  In this case the bottom
quark Yukawa can be large.  However, an examination of the renormalization
group equations shows that this requires a certain amount of fine tuning
(better than $10\%$).
Even if $H_2$ does obtain a vev, it is necessary to
add additional fields to obtain suitable breaking of $SU(2)\times
U(1)$. As is well known, in the MSSM, which only has soft supersymmetry
breaking, if $H_1$ and $H_2$ both obtain negative mass-squared,
the potential is unbounded below. The present case is somewhat
different because not all the supersymmetry breaking terms induced in the
effective theory by radiative corrections are soft, however the
non-supersymmetric dimension four terms are much smaller than the
supersymmetric terms and do not help give an acceptable symmetry breaking.
 Moreover,  in the absence of an $  H_1 H_2  $
coupling, the theory has a Peccei-Quinn symmetry and one obtains a standard
axion. To get around this we add a singlet field, with couplings $$W_S=\tilde
\lambda_1 S H_1 H_2 +  {\tilde\lambda_2\over 3} S^3 \ .\eqn\singletw$$ In order
to understand the absence of other terms,   one can invoke a discrete symmetry.
The terms in \singletw\ gives rise to an effective quartic coupling of the
Higgs
fields, which prevents the runaway behavior.  So one might hope that with this
modification, and with a negative mass-squared only for $H_1$, we could obtain
a sensible breaking of $SU(2) \times U(1)$.

The $  S$ scalar
will obtain a mass at one higher order in the loop expansion than
the Higgs fields, and trilinear terms involving the scalar are also of higher
order.  So, to get a feeling for what may happen, we simply examine the
potential $$\eqalign{V= &-m^2 \vert H_1 \vert^2 +
m^{\prime 2} \vert H_2 \vert^2
+ {g^2 \over 8}(H_1^\dagger \tau^a H_1  +
H_2^\dagger\tau^a H_2)^2 +{g^{\prime 2} \over
8}(H_1^\dagger  H_1 -H_2^\dagger H_2)^2\cr
&+ \vert \tilde \lambda_1 \vert^2(\vert H_1 S\vert^2 + \vert H_2 S
\vert^2) + \vert \tilde \lambda_1 \epsilon_{ij}H_{1i} H_{2j} + \tilde \lambda_2
S^2 \vert^2} \eqn\higgspot$$
However, a detailed study of this potential shows that there is always a scalar
field with a mass less than about $40$ GeV.  In fact,
as it stands, this potential posseses a global symmetry which leads
to a massless pseudoscalar. Corrections to the potential, such as the
nonsupersymmetric cubic terms, will break this symmetry, and give the light
pseudoscalar a mass of order a few GeV.

This situation is unacceptable, given the strong LEP
limits on the decay $Z\rightarrow$ scalar $+$ pseudoscalar.  Preliminary
estimates of further radiative effects indicate that these will not help much.
So we need to consider some further modification.  The simplest possibility
seems to be to add a set of vector-like quarks. In order to maintain the
successful grand unification of the $SU(3)\times SU(2)\times U(1)$ coupling
constants, we can also add vector-like leptons.  We take these fields to have
the quantum numbers of a $5$ and $\bar 5$ of ordinary $SU(5)$.  Denote the
corresponding quarks by $D$ and $\bar D$.  These can couple to the $S$ field,
$S
D \bar D$, and the $\bar D$ field can couple to the Higgs field, $H_2$, $H_2 q
\bar D$, where $q$ is an ordinary quark doublet field.  If these additional
couplings are large enough, several things will happen.  First, the $S$ field
will also obtain a large negative mass-squared at one loop.  It can thus obtain
a large expectation value,   giving rise to a   mass for the
fermionic components of the $D$ and $\bar D$ fields (and the corresponding
leptons).  Second, the field $H_2$ could obtain a large negative contribution
to
its mass-squared at one loop. Also,  the large $S$ vev, in conjuction with the
vev of $H_1$, will induce a vev for $H_2$.   So with this modification,
 a sensible breaking of $SU(2) \times U(1)$ can
arise.  The parameter space of this model is quite large, and we will not
attempt a complete exploration here.  However, it is clear that there are
finite ranges of parameters for which a sensible spectrum is obtained, with all
the scalars heavier than the Z. There is generically a light pseudoscalar, with
mass in the GeV range, which is mainly a gauge singlet.
 Note that with this minimal set of
extra fields, QCD is no longer asymptotically free above the scale of the $f$
and $\bar f$ fields. However, it is almost asymptotically free, and does not
hit
its Landau pole until extremely high energy.

One potential problem with this scenario is that the charge $\frac13$ quarks in
$q$ will mix with those in $D$, giving rise to flavor changing neutral currents
(FCNC) involving the $Z$. However $S$  recieves no positive two loop
contribution to its mass squared, and will get a larger vev than $H_{1,2}$.
Thus the $D$ will be    heavier than the weak scale, which greatly suppresses
its mixing with the light $d$ and $s$ quarks.  The  necessary suppression of
the most severe FCNC, $K_L\rightarrow  \mu^+\mu^-$ , is easily achieved
by requiring that all superpotential  couplings of the lightest two families be
smaller than about $10^{-1}$, including couplings of the form  $H_2
q_{1,2} \bar D$. This   assumption is consistent with the small masses for
these two families and is natural, in the sense of 'tHooft \REF\thooft{'tHooft,
Lecture given at 1979 Carg\`ese Summer Institute.}\refmark\thooft, since a
chiral
flavor symmetry is restored in the limit that all   couplings of the lightest
quark flavors  vanish.

\section{Conclusions}

We have seen that it is in fact possible to construct models with
dynamical supersymmetry breaking at relatively low energy.
We have exhibited a model in which:
\item{a.}  $SU(2) \times U(1)$ is properly broken.
\item{b.}  All superpartners have adequate, calculable masses.
\item{c.}  There is enough degeneracy among quark and lepton masses to assure
absence of flavor-changing neutral currents. This occurs naturally as
a result of the accidental flavor symmetry of the gauge interactions.
\item{d.}  There is no new source of $CP$ violation in the low energy
theory, explaining the absence of large particle electric dipole moments.
\item{e.}  There are no dangerous axions or Goldstone bosons.
\item{f.}  All
couplings are small up to very high energies.
\item{g.}  It is still possible to
unify $SU(3)\times SU(2) \times U(1)$.
\item{h.}  The superpotential is the most
general cubic potential  allowed by the gauge symmetries and  is the most
general consistent with a set of (anomalous) discrete symmetries.
\item{i.} The gravitino is light, of order a keV, and hence provides no
cosmological problems\REF\mmy{T. Moroi, H. Murayama, Masahiro Yamaguchi,
Tohuko University preprint TU-424 (1993), and references therein.}\refmark\mmy.

The model we have described should be viewed as an
existence proof.  Probably the most serious drawback
of this particular model is that the potentially dangerous axion only gains
adequate mass if a certain gauge coupling is in a particular range (the lower
limit set by the mass; the upper limit set by the requirement that the gauge
coupling not blow up too soon).  It would be nice to find a more natural model
which does not suffer from this difficulty.

Some features of the model appear to be generic.
First, the squark and slepton masses are, to a good
approximation functions only of their gauge quantum numbers.
Second, the need for additional fields in order to break $SU(2)\times U(1)$ is
almost certainly general.  The choice we have described
here, of an additional singlet as well as a set of vector-like fermions, is the
simplest possibility we have found.

\REF\preskill{ J. Preskill, S. P. Trivedi, F.
Wilczek, M. B. Wise,  Nucl. Phys. {\bf B363} (1991) 207
}
\REF\antecedents{P. Sikivie, Phys. Rev. Lett. {\bf 48} (1982)
1156; Bob Holdom, Phys. Rev. {\bf D28} (1983) 1419.}
\REF\senjanovic{B. Rai and G. Senjanovic, ICTP preprint IC-92-414.}
\REF\choi{K. Choi, D.B. Kaplan, and A.E. Nelson, preprint UCSD/PTH 92-11
(1992), to appear in Nuclear Physics B, M. Dine, R. Leigh, and D. MacIntire,
Phys. Rev. Lett. {\bf 69} (1992) 2030} Finally, we would like to comment on
some cosmological issues and problems with this model.  Perhaps the most
serious potential problem is one of domain walls.  The model possesses several
spontaneously broken discrete symmetries.  Fortunately,  all of them possess
anomalies with respect to one of the strong gauge groups.  The corresponding
domain walls will thus disappear by the mechanism of Preskill, Trivedi,
Wilczek and Wise.\refmark{\preskill}

In general, however, one might expect non-anomalous discrete
symmetries to arise (this occurs, for example, in the model without
the mirrors).  However, it is not clear that the problem is severe.  Indeed,
the clue to a solution lies in the solution of Preskill et al.  These authors
noted that, even if the scale of the spontaneous breaking is $100$'s of GeV,
the tiny lifting of the degeneracy ($12$ orders of magnitude smaller!) by QCD
is enough to cause collapse of the walls, simply because the expansion
of the universe is so slow.  Suppose, then, that one has some
discrete symmetry without anomalies in the low energy theory.
If this symmetry is broken by non-renormalizable terms, this
will lead to a breaking of the degeneracy.  Even if this effect is quite
small, it can be sufficient to get rid of the domain walls.  For example,
dimension five operators with coefficients slightly larger than
$1/M_p$ or dimension six operators associated with a scale of order
$10^9 $ GeV  or smaller should be enough.
This solution to the domain wall problem can be relevant quite
generally, and has antecedents\refmark{\antecedents}
in earlier work on axions and
technicolor. \foot{ Recently it has been discussed for spontaneous breaking of
$P$ and $CP$ symmetries.\refmark{\senjanovic}
However, in the cases which
have been studied, there are low dimension operators in the low
energy theory which can break the symmetry.  Moreover, the most
plausible context for spontaneous breaking of these symmetries is
in theories in which these symmetries are gauge symmetries, in
which case there is no explicit breaking of the symmetry.\refmark{\choi}}

In these models, one must also study the possibility of stable or nearly stable
massive particles, such as the ``$f$'' fermions.  Again, it may be
necessary to invoke higher dimension operators to allow these to decay
and avoid cosmological problems.  This problems  may not be generic, but
 specific to the model  under study.

In any case, we believe that models with low energy dynamical
supersymmetry breaking in the visible sector are a plausible alternative to
more conventional hidden sector supergravity models.  They
solve some of the most troubling problems of the hidden sector models,
and they provide a dynamical solution of the hierarchy problem.

\bigskip\bigskip \centerline{{\bf Acknowledgements}} The
work of A.N.    was supported in part by the Department of
Energy under
contract \#DE-FGO3-90ER40546, the Alfred P. Sloan
Foundation and the Texas National Laboratory Research Commission. The work of
M.D. was supported in part by the DOE under contract \#DE-FG03-92ER40689.

\figout
\refout
\end
\vfill
\epsfbox{figone.eps}
Figure 1
\par
\vfill
\epsfbox{figtwo.eps}
Figure 2
\vfill
\eject
\vfill
\epsfbox{figthree.eps}
Figure 3
\par
\vfill
\epsfbox{figfour.eps}
Figure 4
\eject
\vfill
\epsfbox{figfive.eps}
Figure 5
\par
\vfill
\epsfbox{figsix.eps}
Figure 6
\vfill
\eject
\vfill
\epsfbox{figseven.eps}
Figure 7
\par
\vfill
\epsfbox{figeight.eps}
Figure 8
\vfill
\eject
\end